\documentclass[11pt,a4paper]{article}
\pdfoutput=1
\usepackage{jheppub}

\usepackage{amssymb,amsmath}
\usepackage{amsfonts}
\usepackage{mathtools}

\let\originalleft\left
\let\originalright\right
\renewcommand{\left}{\mathopen{}\mathclose\bgroup\originalleft}
\renewcommand{\right}{\aftergroup\egroup\originalright}

\makeatletter
\newenvironment{equations}[1][]{\subequations\ifx\relax#1\relax\else\label{#1}\fi\align\ignorespaces}{\endalign\ignorespacesafterend\endsubequations}
\def\@spliteq#1{\begin{equation}\begin{split}#1\end{split}\end{equation}}
\def\splitequation{\collect@body\@spliteq}

\makeatother

\newcommand{\eqend}[1]{\,#1}

\usepackage{xspace}
\renewcommand{\d}{\ifmmode\operatorname{d}\!\else\textrm{d}\xspace\fi}
\newcommand{\ig}{{i}_\gamma}
\newcommand{\UR}{\mathrm{U}(1){}_\text{R}}
\newcommand{\Tr}{\operatorname{Tr}}

\newcommand{\mathi}{\mathrm{i}}
\newcommand{\abs}[1]{{\left\lvert{#1}\right\rvert}}

\makeatletter
\gdef\@fpheader{}
\makeatother

\title{Deformations of Supergravity and Supersymmetry Anomalies}

\author[a]{Markus B. Fr\"ob,}
\author[b,c]{Camillo Imbimbo}
\author[d,e]{and Nicol\`o Risso}

\affiliation[a]{Institut f\"ur Theoretische Physik, Universit\"at Leipzig, Br\"uderstra\ss e 16, 04103 Leipzig, Germany}
\affiliation[b]{Dipartimento di Fisica, Universit\`a di Genova, Via Dodecaneso 33, 16146 Genoa, Italy}
\affiliation[c]{INFN, Sezione di Genova, Genoa, Italy}
\affiliation[d]{Dipartimento di Fisica e Astronomia ``Galileo Galilei'', Universit\`a di Padova, Via F. Marzolo 8, 35131 Padua, Italy}
\affiliation[e]{INFN, Sezione di Padova, Padua, Italy}

\emailAdd{mfroeb@itp.uni-leipzig.de}
\emailAdd{camillo.imbimbo@ge.infn.it}
\emailAdd{nicolo.risso@pd.infn.it}

\abstract{We present a BRST analysis of supersymmetry anomalies of $\mathcal{N} = 1$ supersymmetric quantum field theories with anomalous R symmetry. To this end, we consider the coupling of the matter theory to classical $\mathcal{N} = 1$ new minimal supergravity. We point out that a supersymmetry anomaly cocycle associated to the $\UR$ field does exist for this theory. It is non-trivial in the space of supergravity fields (and ghosts), but it becomes BRST-exact in the functional space that includes antifields. Equivalently, the $\UR$ supersymmetry anomaly cocycle vanishes ``on-shell''. It is therefore removable. However, to remove it ---  precisely because it is not trivial in the smaller space of fields ---  one needs to deform the supergravity BRST operator. This deformation is triggered, at first order in the anomaly coefficient, by a local operator $S_1$ of ghost number 1. We give a cohomological characterization of $S_1$ and compute it in full detail. At higher orders in the anomaly coefficient, we expect a priori that further deformations of the BRST rules are necessary.}


\begin{document}

\maketitle

\section{Introduction and summary}

Classical symmetries can be anomalous. The possibility that supersymmetry is affected by quantum anomalies has been investigated, with different methods, for a long time~\cite{Piguet:1980fa,Nielsen:1984nk, Itoyama:1985qi, Guadagnini:1985ea, Bonora:1984pn,Bonora:1985ug, Ferrara:1985me, Brandt:1996au}. More recently this question has attracted renewed interest, starting from some explicit computations that described anomalous supersymmetric Ward identities for 4-dimensional supersymmetric matter quantum field theories (SQFT's) with anomalous R symmetry~\cite{Papadimitriou:2017kzw, Papadimitriou:2019yug, Katsianis:2019hhg, Katsianis:2020hzd, An:2019zok}. The relevance of these computations for supersymmetry anomalies has been discussed in~\cite{Kuzenko:2019vvi,Bzowski:2020tue,Minasian:2021png}.

In this work we reconsider this question in the BRST framework. In this setup one couples a generic matter SQFT to a \emph{classical} supergravity background, whose local supersymmetry transformations close off-shell. Each supergravity field is the classical source of some matter quantum current. For each symmetry of the theory, one then introduces a ghost field with opposite statistics. The nilpotent BRST operator is given by the sum of all symmetry transformations (with the ghost fields replacing the transformation parameters) and acts on both background supergravity fields and ghosts. In this framework, anomalies are elements of the non-trivial cohomology class of ghost number 1 of the BRST operator (i.e., cocyles of ghost number 1), and are local functionals of the supergravity fields and ghosts. The matter SQFT enters the analysis only via its global symmetries which specify the background supergravity BRST rules.

Since our goal is to analyze super-anomalies of SQFT's with R symmetries, we will pick $\mathcal{N} = 1$ new minimal supergravity~\cite{Sohnius:1981tp,Ferrara:1988qxa} as the background supergravity theory. The multiplet of $\mathcal{N} = 1$ new minimal supergravity includes the abelian vector field gauging the $\UR$ symmetry. When the matter SQFT has other global (flavor) symmetries beyond R symmetries, one can study the corresponding anomalies by coupling $\mathcal{N} = 1$ new minimal supergravity to background super-Yang--Mills (super-YM) multiplets whose fields source the flavor currents and their superpartners. In this paper we are interested in the supersymmetric anomalies associated to R symmetries, and thus most of our analysis will focus on pure $\mathcal{N} = 1$ new minimal supergravity. Nevertheless, the discussion in Sections~\ref{sec:sugrabrst} and~\ref{sec:susyanomalyrev} is general and applies to any supergravity theory.

The local BRST cohomology of $\mathcal{N} = 1$ new minimal supergravity has been investigated in the past. As far as we know, the most complete classification of its local cocycles is found in~\cite{Brandt:1996au}. In that work, a restricted number of anomaly BRST cocycles are listed, among which there is the celebrated supersymmetrization of bosonic YM anomalies, previously discovered  in~\cite{Bonora:1985ug, Itoyama:1985qi}. We will refer to this BRST cocycle as the supersymmetric chiral anomaly cocycle. The other cocycles listed in~\cite{Brandt:1996au} do not seem to be relevant
to the questions raised by the works~\cite{Papadimitriou:2017kzw, Papadimitriou:2019yug, Katsianis:2019hhg, Katsianis:2020hzd}.

The analysis of~\cite{Brandt:1996au} employs methods which are very general and powerful, but also quite abstract. To make this paper self-contained, we present in Section~\ref{sec:susyanomalyrev} a (to the best of our knowledge) novel derivation of the supersymmetric chiral anomaly cocycle which is both very simple and geometric in character, and closely parallels the classic derivation of the bosonic YM anomaly cocycles of~\cite{Bardeen:1984pm,Manes:1985df}. In addition, this derivation also makes it very clear, in a completely model-independent way, that there exists a supersymmetric chiral anomaly cocycle for \emph{each} of the YM symmetries of the background supergravity. In particular, this BRST cocycle exists also for the $\mathrm{U}(1)$ R symmetry whose gauge field sits in the $\mathcal{N} = 1$ new minimal supergravity multiplet.

There is, however, an important and crucial difference between the $\UR$ super-cocycle and those associated to global ``flavor'' symmetries of the matter SQFT --- YM symmetries whose gauge fields lie in external super-YM multiplets coupled to supergravity. Although both kind of cocycles are non-trivial in the space of supergravity (and ghosts) fields, the supersymmetric cocycle associated to the $\mathrm{U}(1)$ R symmetry becomes trivial after one extends the BRST operator to the larger space of functionals of both supergravity fields and their \emph{antifields}.\footnote{For this reason, the $\UR$ super-cocycle is not listed among the non-trivial anomaly cocycles in~\cite{Brandt:1996au}. In this paper it was already observed that the cocycle in question is trivial in the full BV formalism.} This is not the case for the anomaly super-cocycle associated to a ``flavor'' $\mathrm{U}(1)$ super-YM multiplet, which remains non-trivial and cannot be removed even after the introduction of antifields.

It should be kept in mind that the transformation rules of $\mathcal{N} = 1$ new minimal supergravity close off-shell, and therefore its BRST operator is already nilpotent on the space of fields (including the ghosts), without the need to introduce antifields. Since its cohomology problem is therefore perfectly well-defined on the space of fields and ghosts only, the non-triviality of the $\UR$ super-anomaly cocycle on this space is a mathematically well-defined concept.

When one considers a dynamical theory one expects that radiative corrections may, in general, renormalize not just the action but also the BRST transformations. For this reason, the general consensus is that anomalies which trivialize when one introduces antifields are not ``true'' anomalies (at least for a dynamical theory). Indeed, one can convince oneself that this kind of anomaly cocycles can be removed by adding local counterterms not just to the local action, but also to the BRST transformations. Nevertheless, to actually remove such anomalies in practice, one needs to know how to \emph{explicitly} write down the anomaly cocycle as the BRST variation of a local functional of both fields and antifields: it is this functional that contains the information about the renormalization of both the action and the BRST rules. To our knowledge, the fully explicit trivialization of the $\UR$ super-cocycle of $\mathcal{N} = 1$ new minimal supergravity has not been determined yet --- neither in~\cite{Brandt:1996au}\footnote{In eq. (6.10) of \cite{Brandt:1996au} some parts of the trivialization of the anomaly super-coycle in the BV framework are given.} nor in following works. What we do in the present article amounts to solving this problem.

Moreover, when one is dealing with a theory of \emph{classical sources} whose local symmetries close off-shell, as we are doing here, one might be skeptical about the actual necessity of introducing antifields, which are sources for the BRST variations of the (classical) sources. In other words, one may wonder if it is possible --- and maybe more practical --- to characterize
anomaly cocycles which are non-trivial in field space but become trivial in the enlarged space of fields and antifields without resorting to the full power of the antifield (Batalin--Vilkovisky, BV) formalism. In Section~\ref{sec:evanescent anomalies} we will do precisely that: We will describe these kind of anomalies and their removal in the standard BRST framework, without introducing antifields.

The anomalies in question vanish ``on-shell'', i.e., they are proportional to the ``equations of motion'' of some BRST-invariant functional of the fields --- for example the classical supergravity action. For lack of a better name, we will call these anomalies \emph{evanescent}. However, it is worth reiterating that we are considering here a theory of classical, non-dynamical sources\footnote{Being non-renormalizable, new minimal $\mathcal{N} = 1$ supergravity is of course not believed to be a consistent as a dynamical theory at the quantum level. The fact that its anomalies are removable leaves open the possibility that it admits a consistent ultraviolet completion.}: therefore the expression ``equations of motion'' does not refer to any kind of dynamics, it is merely a shorthand for functional derivatives of local functionals.

We will explain that by writing down evanescent anomalies in terms of the equations of motion, one can read off candidate deformations of the original BRST transformations, up to certain ambiguities which one can try to resolve by imposing that the deformation \emph{anticommutes} with the original BRST transformations. If this can be achieved, the \emph{deformed} BRST transformations leave a \emph{deformed} effective action invariant (to first order in the anomaly coefficient): the latter one equals the sum of the original ---  non-local and anomalous ---  effective action and of the local invariant action which defines the equations of motion.  This double deformation --- both of the action and of the transformation laws --- removes the original anomaly super-cocycle, and the resulting theory has no anomalies at all (to first order) --- neither supersymmetry anomalies nor $\UR$ gauge anomalies.

To extend this procedure to all orders in the coefficient of the anomaly, certain integrability conditions have to be met, which we will spell out in detail in Section~\ref{sec:evanescent anomalies}.  In short, the anomaly can be removed if there is no operator of ghost number higher than 1 which (anti-)commutes with the original BRST operator and which is not an (anti-)commutator of the original BRST operator with some functional derivative. In other words, a sufficient condition for the removal of the anomaly is the vanishing of higher ghost number cohomologies of the original BRST operator on the space of (local) functional derivatives, on which the BRST operator acts by (anti-)commutators. The deformation at first order in the anomaly coefficient is instead a non-trivial element of the same cohomology at ghost number 1.

These same conditions can be phrased in the language of antifields. In the BV formalism, deformations of the BRST operator which remove the anomaly are associated to the cohomology at ghost number 0 of the original BRST operator, acting now on the space of functionals of fields and antifields, and obstructions to integrate them are associated to cohomologies at higher ghost number. However, there is an important caveat that distinguishes the problem of anomaly removal, which we are describing here, from the superficially similar one of finding the deformations of the BV master equation, as described for example in~\cite{Barnich:1993vg}. In the latter context, the relevant concept is the one of \emph{local} cohomologies on the space of functionals of fields and antifields. In the supergravity case, it was shown in~\cite{Brandt:1996au} that the deformations that change the symmetry transformations always involve other external supermultiplets in addition to the supergravity one. On the other hand, deformations of the BRST operator which remove evanescent anomalies are associated to the effective action obtained by integrating out some SQFT. This effective action is a \emph{non-local} functional of the supergravity fields. Therefore the results in~\cite{Brandt:1996au} regarding the local cohomology at ghost number 0 of new minimal $\mathcal{N} = 1$ supergravity do not contradict the existence of a deformation of the BRST operator removing the $\UR$ supersymmetric anomaly.

We will compute explicitly such a deformation, at first order in the anomaly coefficient, in Section~\ref{sec:deformation}: the deformation is given in Eqs.~\eqref{FinalS1}, and represents our most important new result. It turns out that to remove the anomaly one needs to deform the BRST transformations of both the gravitino and the auxiliary fields of $\mathcal{N} = 1$ new minimal supergravity, which are the two-index antisymmetric gauge field $B_{\mu\nu}$ and the $\UR$ gauge field. Since supergravity is non-renormalizable, the deformation is described by operators of higher dimension. For this reason we also expect that the deformation necessary to remove the anomaly at all orders in the anomaly cofficient will include terms of all orders in the derivative
expansion. We leave to the future the problem of computing higher order corrections (and possible obstructions) to the BRST transformations necessary to remove the evanescent $\UR$ superanomaly.

The rest of the article is organized as follows: In Section~\ref{sec:sugrabrst} we review the BRST formulation of supergravity and the notion of \emph{equivariant} BRST operator~\cite{Imbimbo:2018duh}, which we employ instead of the full BRST operator to keep the computations manageable. This BRST operator, equivariant with respect to diffeomorphisms and to the other bosonic gauge symmetries of supergravity, involves only the commuting ghost of local supersymmetry $\zeta$. Nevertheless, the cohomology of the standard nilpotent BRST supergravity operator on forms modulo the exterior differential \d, i.e., the cohomology of the standard BRST operator acting on integrated local functionals, is completely equivalent to the cohomology of the equivariant BRST operator modulo both \d and $\ig$. Here, $\ig$ is the nilpotent operator that contracts forms with the commuting vector field $\gamma^\mu$, the universal bilinear of the supersymmetry ghost $\zeta$ given in Eq.~\eqref{gammazetazeta}.

In Section~\ref{sec:susyanomalyrev} we present a derivation of the supersymmetric chiral anomaly cocycle which involves the super-Chern classes built out of the super-connection, the sum of YM gauge fields and their ghosts that was introduced in~\cite{Manes:1985df}. The crucial difference between supergravity and bosonic gauge theories is that the supergravity super-Chern classes have non-vanishing components of higher ghost number. For this reason, one cannot simply identify the supergravity anomaly cocycle with the Chern-Simons super-form, as one does in the bosonic case. We will explain that when the component of the super-Chern class of higher ghost number is $\ig$-trivial --- as it is the case for $\mathcal{N} = 1$ new minimal supergravity --- it is nevertheless possible to complete the Chern-Simons superform to produce a BRST cocycle, the supersymmetric chiral anomaly. We will show how the existence of this cocycle can be neatly understood in terms of the cohomology of $\ig$.
 
In Section~\ref{sec:Bsector} we describe the (possibly not very familiar) equivariant BRST structure of the antisymmetric gauge field $B_{\mu\nu}$ sector of $\mathcal{N} = 1$ new minimal supergravity.

In Section~\ref{sec:newN1BRSTrules} we present the BRST rules of the model and the form of the $\UR$ supersymmetric anomaly cocycle.

Section~\ref{sec:evanescent anomalies} contains a general discussion of ``evanescent'' BRST anomaly cocycles in the framework which does not involve antifields. We spell out the conditions for these cocycle to be removable.

In Section~\ref{sec:deformation} we finally evaluate the deformation of the BRST operator of $\mathcal{N} = 1$ new minimal supergravity which removes the supersymmetric $\UR$ anomaly at first order in the coefficient of the anomaly, using the \textsc{FieldsX} extension package~\cite{fieldsx} for the \textsc{xAct} tensor algebra suite~\cite{xact}.

In Section~\ref{sec:renormalizedtorsion} we discuss how the deformation affects the constraints connecting the super-torsion to the super-Chern classes of the Lorentz and $\UR$ local symmetries.
 
In the conclusions, Section~\ref{sec:conclusions}, we summarize our findings and briefly discuss their relation with the recent works~\cite{Papadimitriou:2019yug,Kuzenko:2019vvi,Bzowski:2020tue} on the same topic. 

Lastly, in Appendix~\ref{sec: relation with BV} the integrability conditions for $S_1$ that were described in Section \ref{sec:evanescent anomalies} are reformulated in the BV language involving antifields, and in Appendix~\ref{sec:fieldsx} we give some details on the computation using \textsc{FieldsX}.

\section{The equivariant BRST operator of supergravity}
\label{sec:sugrabrst}

In the BRST framework one introduces ghost fields of ghost number +1 in correspondence to each of the local symmetries. Among the \emph{bosonic} local symmetries of supergravity there are diffeomorphisms and YM gauge symmetries. $\mathcal{N} = 1$  new minimal supergravity is also invariant under local vector-like gauge transformations whose gauge field is an antisymmetric tensor $B_{\mu\nu}$. We will postpone dealing with those to Section \ref{sec:Bsector}: they will modify the algebraic structures that we will describe in this Section in some relatively obvious way. We will denote by $\xi^\mu$ the \emph{anticommuting} vector ghost field associated to diffeomorphisms, and by $c$ the \emph{anticommuting} ghost associated to the YM gauge symmetry which takes values in the adjoint representation of the YM algebra. The YM gauge symmetries always include local Lorentz transformations. Beyond local Lorentz gauge symmetry, we will also allow for additional YM gauge symmetries: among those the one corresponding to the R symmetries of the SQFT whose coupling to supergravity one is considering.

In correspondence with $\mathcal{N}$ local supersymmetries, one introduces \emph{commuting} supergravity spinorial Majorana ghosts $\zeta^i$ with $i=1, \ldots, \mathcal{N}$, whose BRST transformation rules have the form
\begin{equation}
s\, \zeta^i = \ig(\psi^i) + \text{diffeos} + \text{gauge transformations} \eqend{.}
\label{szeta}
\end{equation}
In this equation $s$ is the \emph{nilpotent} BRST operator
\begin{equation}
s^2 = 0 \eqend{,}
\end{equation}
$\psi^i = \psi^i_\mu \d x^\mu$ are the Majorana gravitinos, and $\gamma^\mu$ is the following vector bilinear of the commuting ghosts\footnote{To avoid confusions, we will denote with $\gamma^\mu$ the ghost bilinear and with $\Gamma^a$ the Dirac matrices.} 
\begin{equation}
\gamma^\mu \equiv \sum_i \bar\zeta^i \, \Gamma^a \, \zeta^i \, e_a{}^\mu \eqend{,}
\label{gammazetazeta}
\end{equation}
where $e_a{}^\mu$ are the inverse of the vierbein $e^a \equiv e^a{}_\mu \d x^\mu$. The vector $\gamma^\mu$ has ghost number +2. Both the Majorana ghosts $\zeta^i$ and gravitinos $\psi^i$ carry a label $i=1,\ldots, \mathcal{N}$ on which the $\mathrm{O}(\mathcal{N})$ subgroup of the R symmetry group acts. However, the full R symmetry group can be as large as $\mathrm{U}(\mathcal{N})$. In the following we will restrict ourselves to the $\mathcal{N} = 1$ case, and consequently omit the index $i$.

The BRST transformations of the vierbein are universal, i.e., valid for any supergravity theory:
\begin{equation}
s\, e^a = - \bar\zeta \, \Gamma^a \, \psi + \text{diffeos} + \text{gauge transformations} \eqend{.}
\label{svier}
\end{equation}
We will denote the action of diffeomorphisms with $\mathcal{L}_\xi$, the Lie derivative associated with the vector field $\xi^\mu$. Let us also denote the YM gauge transformations with odd parameter $c$ by $\delta_c^\text{\tiny YM}$. The BRST transformations of the diffeomorphism ghost are
\begin{equation}
s \, \xi^\mu = - \frac{1}{2} \mathcal{L}_\xi \xi^\mu + \gamma^\mu \eqend{.}
\label{sxi}
\end{equation}
The BRST transformations~\eqref{szeta} and~\eqref{svier} imply~\cite{Baulieu:1985md} that the transformation rule for the vector ghost bilinear $\gamma^\mu$ is also universal:
\begin{equation}
s \, \gamma^\mu = - \mathcal{L}_\xi \gamma^\mu \eqend{.}
\label{sgamma}
\end{equation}
It is then convenient to introduce the fermionic ghost number 1 operator
\begin{equation}
S \equiv s + \mathcal{L}_\xi + \delta^\text{\tiny YM}_c \eqend{,}
\end{equation}
which we will call the \emph{equivariant} BRST operator. $S$ is not nilpotent: Nilpotency of the full BRST operator $s$ is equivalent to the relation~\cite{Imbimbo:2018duh}
\begin{equation}
S^2 = \mathcal{L}_\gamma + \delta_{ \ig(A) + \phi}
\label{Salgebra}
\end{equation}
where $\gamma^\mu$ is defined in Eq.~\eqref{gammazetazeta},
\begin{equation}
\ig(A) \equiv \gamma^\mu A_\mu
\end{equation}
is the contraction of the YM field 1-form with the vector $\gamma^\mu$, and $\phi$ is a bilinear of $\zeta$ with values in the YM gauge algebra which we will comment on in a moment.

It is important to note that the equivariant algebra~\eqref{Salgebra} holds on all fields except the anticommuting ghosts $\xi^\mu$ and $c$ associated to the bosonic local symmetries. On the ghosts $\xi^\mu$ and $c$, only the action of the nilpotent $s$ --- but not that of the equivariant $S$ --- is meaningfully defined. The action of $s$ on the diffeomorphism ghost $\xi^\mu$ has been written above in Eq.~\eqref{sxi};  the BRST transformation of the gauge ghost $c$ turns out to involve the bilinear $\phi$ that appears in the equivariant algebra~\eqref{Salgebra}:
\begin{equation}
s\, c = - c^2 + \mathcal{L}_\xi c + \ig(A) + \phi \eqend{.}
\label{scYM}
\end{equation}

The relevance of the equivariant BRST operator $S$ is as follows: the cohomology of the nilpotent BRST operator $s$ modulo the exterior differential \d on local forms that depend on both fields and ghosts $\xi$, $c$ and $\zeta$, i.e., on integrated local functions of fields and ghosts --- the object of interest in local quantum field theories --- is isomorphic to the cohomology of the equivariant $S$ modulo both \d and $\ig$ on invariant forms which depend on the fields and $\zeta$. In other words, the benefit of introducing $S$ is to work on the smaller space of fields and $\zeta$, forgetting about $c$ and $\xi$, which greatly simplifies the analysis.

The universal BRST transformation rules~\eqref{sgamma} imply that $\gamma^\mu$ is $S$-invariant:
\begin{equation}
S\, \gamma^\mu = 0 \eqend{.}
\label{gammaconsistency}
\end{equation}
The bilinear $\phi$, however, is model-dependent: Different supergravity theories are characterized by different $\phi$'s. In general, we can only assert that $\phi$ must satisfy a consistency condition which comes from the nilpotency of $s$:
\begin{equation}
S\, \phi = \ig(\lambda) \eqend{,}
\label{phiconsistency}
\end{equation}
where $\lambda$ is the $S$-variation of the gauge field
\begin{equation}
S\, A \equiv \lambda \eqend{,}
\end{equation}
a one-form of ghost number 1 with values in the YM algebra, which will be referred to as the \emph{topological gaugino}. The equivariant relation~\eqref{Salgebra} implies that $A$, $\lambda$ and $\phi$ all sit in a BRST multiplet with values in the adjoint of the gauge algebra:
\begin{equations}[Sgauge]
S\, A &= \lambda \eqend{,} \\
S\, \lambda &= \ig(F) - D\, \phi = \ig(F) - \d \phi - [A,\phi] \eqend{,} \\
S\, \phi &= \ig(\lambda) \eqend{.}
\end{equations}
These relations are again universal, i.e., valid for generic supergravities, and only the concrete form of $\phi$ depends on the specific theory. They are structurally identical to the BRST rules for topological YM theory coupled to topological gravity~\cite{Imbimbo:2014pla,Imbimbo:2018duh}.

\section{The supersymmetry anomaly revisited}
\label{sec:susyanomalyrev}
The equivariant BRST framework for generic $\mathcal{N} = 1$ supergravity just reviewed in the previous Section allows for a transparent and geometric description of the celebrated supersymmetric chiral anomaly BRST cocycle.

Anomalies are best described by introducing generalized forms (or ``super-forms'') with fixed \emph{total fermionic} degree, which is the sum of form and ghost number degrees. For example, the generalized connection defined as
\begin{equation}
\mathbb{A} \equiv c + A
\label{superconnection}
\end{equation}
is a generalized form of total fermionic degree 1. Let us also define a coboundary operator which extends the BRST action to generalized forms: for supergravity theories this is
\begin{equation}
\delta \equiv s + \d{} + \mathcal{L}_\xi - \ig = S - \delta^\text{\tiny YM}_c + \d{} - \ig \eqend{.}
\label{deltacoboundary}
\end{equation}
The equivariant BRST algebra~\eqref{Salgebra} ensures that $\delta$ is nilpotent:
\begin{equation}
\delta^2 = 0 \eqend{.}
\end{equation}
On then defines the generalized curvature of the super-connection~\eqref{superconnection}
\begin{equation}
\mathbb{F} \equiv \delta\, \mathbb{A} + \mathbb{A}^2 = F + \lambda + \phi \eqend{,}
\label{superF}
\end{equation}
which satisfies the generalized or super-Bianchi identity
\begin{equation}
\delta \,\mathbb{F} + [\mathbb{A}, \mathbb{F}] = 0 \eqend{.}
\label{superBianchi}
\end{equation}
It is important to remark that the same identical construction also works for the BRST formulation of non-supersymmetric gauge theories. In this case however, the generalized curvature and the ordinary curvature coincide, $\mathbb{F} = F$, which follows from Eq.~\eqref{superF} by putting the supersymmetry ghost $\zeta$ to zero such that $\lambda = 0 = \phi$.

The super-Bianchi identity~\eqref{superBianchi} implies that the super-Chern classes built with $\mathbb{F}$ are $\delta$ cocycles, and in particular one has
\begin{equation}
\delta \Tr \mathbb{F}^3 = 0 \eqend{.}
\label{superChernclosed}
\end{equation}
One shows also in the standard way that such classes are $\delta$-exact:
\begin{equation}
\Tr \mathbb{F}^3 = \delta\, \Gamma_5(\mathbb{A}, \mathbb{F}) \eqend{,}
\label{superCS}
\end{equation}
where $\Gamma_5(\mathbb{A}, \mathbb{F})$, the celebrated Chern--Simons functional, is a polynomial in $\mathbb{A}$ and $\mathbb{F}$. It is a generalized form of total fermionic degree 5:
\begin{equation}
\Gamma_5(\mathbb{A}, \mathbb{F}) = \Tr \Bigl[ \mathbb{F}^2 \mathbb{A} - \frac{1}{2} \mathbb{F} \mathbb{A}^3 - \frac{1}{10} \mathbb{A}^5 \Bigr] \eqend{.}
\label{superCSbis}
\end{equation}
The relevance of this construction to anomalies is as follows: The anomaly is the BRST variation of the effective action, and consequently its BRST variation vanishes.\footnote{This is the analog of the Wess--Zumino consistency condition in the BRST formalism.} In 4 dimensions, the anomaly can therefore be obtained from a generalized form of total fermionic degree 5 which is a $\delta$ cocycle.

To show how one can obtain the anomaly from such a generalized form, let us first consider non-supersymmetric \emph{bosonic} YM gauge (and local Lorentz) symmetries. As mentioned above, in this case  we have
\begin{equation}
\mathbb{F} = F \eqend{.}
\label{BosonicFFbb}
\end{equation}
Since $F$ is an ordinary two-form, for \emph{bosonic} gauge symmetries the super-Chern class of degree 6~\eqref{superChernclosed} is an ordinary 6-form and vanishes in 4 dimensions:
\begin{equation}
\Tr \mathbb{F}^3 = \Tr F^3 = 0 \eqend{.}
\label{YManomalies}
\end{equation}
It follows from Eq.~\eqref{superCS} that the Chern--Simons super form $\Gamma_5(\mathbb{A},\mathbb{F})$ is $\delta$-closed:
\begin{equation}
\delta \, \Gamma_5(\mathbb{A}, \mathbb{F}) = 0 \eqend{,}
\end{equation}
Hence the 4-form component of $\Gamma_5(\mathbb{A}, \mathbb{F})$, which has ghost number 1, is $s$-closed modulo $\d{}$: it is the anomaly of YM gauge and local Lorentz symmetries. Its explicit form is readily obtained from~\eqref{superCSbis} by inserting the definition~\eqref{superconnection} for the generalized connection $\mathbb{A}$.

In the supergravity case this story requires modifications. Indeed, as can be seen in Eq.~\eqref{superF}, the super-curvature $\mathbb{F}$ of supergravity has both 1-form and 0-form components $\lambda$ and $\phi$ which, in general, do not vanish. Therefore the Chern class $\Tr \mathbb{F}^3$ does not vanish in 4 dimensions, and the super Chern--Simons functional is not $\delta$-closed and thus not a cocycle. However, not all is lost. To start with, let us remark that $S$ and $\ig$ anticommute
\begin{equation}
\{ S, \ig \} = 0
\end{equation}
thanks to Eq.~\eqref{gammaconsistency}. Moreover, the BRST variation of the 0-form component $\phi$ of the super-curvature $\mathbb{F}$ is always $\ig$-exact~\eqref{phiconsistency}. Now, \emph{in a certain class} of supergravity theories, $\phi$ \emph{itself} is $\ig$-exact:
\begin{equation}
\phi = - \ig (H) \eqend{,}
\end{equation}
where $H$ is a 1-form of ghost number 0. For these theories one can define a new connection
\begin{equation}
A_- \equiv A - H
\label{Aminusconnection}
\end{equation}
and rewrite the equivariant BRST algebra~\eqref{Salgebra} with a vanishing $\phi$:
\begin{equation}
S^2 = \mathcal{L}_\gamma + \delta_{ \ig(A_-) } \eqend{.}
\label{Salgebraminus}
\end{equation}
Hence the super curvature $\mathbb{F}_-$ associated to the super-connection
\begin{equation}
\mathbb{A}_- \equiv c + A_-
\label{superconnectionminus}
\end{equation}
has vanishing 0-form component
\begin{equation}
\mathbb{F}_- = F_- + \lambda_- \eqend{,}
\end{equation}
where
\begin{equations}
F_- &\equiv \d A_- + A_-^2 \eqend{,} \\
\lambda_- &\equiv S\, A_- \eqend{.} \label{lambdaminusfromsaminus}
\end{equations}
Moreover, the equivariant BRST algebra~\eqref{Sgauge} ensures that the topological gaugino associated to the connection $A_-$ is $\ig$-closed:
\begin{equation}
\ig(\lambda_-) = 0 \eqend{.}
\label{lambdaclosed}
\end{equation}
The super-Chern class built with $\mathbb{F}_- $ still does not vanish in 4 dimension, but we have
\begin{equation}
\Tr \mathbb{F}_-^3 = 3 \Tr F_- \, \lambda_-^2 + \Tr \lambda_-^3 \eqend{.}
\end{equation}
Its $\delta$-closedness~\eqref{superChernclosed} implies that its component with lowest form degree is $\ig$-closed:
\begin{equation}
\ig \bigl( \Tr \lambda_-^3 ) = 0 \eqend{,}
\end{equation}
which indeed follows directly from Eq.~\eqref{lambdaclosed}. As we have already said, $\ig$ is nilpotent:
\begin{equation}
\ig^2 = 0 \eqend{,}
\end{equation}
and it is therefore sensible --- and useful --- to consider the cohomology of $\ig$ on the space of generalized gauge-invariant forms. The crucial question, to understand supersymmetric anomalies, is to establish if the \emph{gauge-invariant}, $\ig$-closed 3-form $\Tr \lambda_-^3$ of ghost number 3 is a trivial element of the $\ig$-cohomology on \emph{gauge-invariant} forms. Suppose, for the moment, that this is the case:
\begin{equation}
\Tr \lambda_-^3 = - \ig\big( \Omega^{(4)}_1 \big) \eqend{,}
\end{equation}
where $\Omega^{(4)}_1$ is a \emph{gauge-invariant} 4-form of ghost number 1. It then follows that the 4-form component of the super Chern class is $S$-exact:
\begin{equation}
3 \Tr F_- \, \lambda_-^2 = S\, \Omega^{(4)}_1 \eqend{,}
\end{equation}
and that the super-Chern class is $\delta$-exact
\begin{equation}
\Tr \mathbb{F}_-^3 = \delta \, \Omega^{(4)}_1 \eqend{.}
\end{equation}
Plugging this back into Eq.~\eqref{superCS}, we obtain
\begin{equation}
\delta \left[ \Gamma_5(\mathbb{A}, \mathbb{F}) - \Omega^{(4)}_1 \right] = 0 \eqend{.}
\end{equation}
Hence the candidate supersymmetry anomaly is
\begin{equation}
\mathcal{A}_5 = \Gamma_5(\mathbb{A}, \mathbb{F}) - \Omega^{(4)}_1 \eqend{,}
\label{superCSanomaly}
\end{equation}
which is indeed the ``supersymmetrization'' of the bosonic gauge anomaly.

Summarizing, if the invariant $\Tr \lambda_-^3$ is $\ig$-exact, then the supersymmetric extension \eqref{superCSanomaly} of the Chern--Simons functional is an anomaly supergravity cocycle. Let us therefore discuss briefly the structure of the $\ig$-cohomology.

The topological gaugino $\lambda_-$ has ghost number 1, and thus it is certainly not $\ig$-exact on the space of fields, which does not include negative ghost numbers. It is however $\ig$-closed thanks to the Fierz identity involving 3 supersymmetry Majorana ghosts:
\begin{equation}
\big( \bar\zeta\, \Gamma^\mu \zeta \big) \bar\zeta\, \Gamma_\mu = 0 \eqend{,}
\label{zetafierz3}
\end{equation}
where we recall that we denote the (4-dimensional) Dirac matrices by $\Gamma^\mu$. The general solution of the consistency condition~\eqref{lambdaclosed} has thus the form
\begin{equation}
\lambda_- = \bar\zeta\, \Gamma_\mu \chi \d x^\mu \eqend{,}
\label{topgauginostructure}
\end{equation}
where $\chi$ is a Majorana spinor field of ghost number 0 with values in the YM gauge Lie algebra. We will call $\chi$ the ``spinorial gaugino'', and generally it might or might not be an elementary field. If $A$ belongs to a super-Yang--Mills multiplet, $\chi$ is precisely the usual ``physical'' gaugino. However, if $A$ belongs to the supergravity multiplet, then $\chi$ is a composite field. For example, $\mathcal{N} = 1$ new minimal supergravity contains both a $\UR$ axial gauge field $A_-$ which gauges the R symmetry and a $\mathrm{SO}(1,3)$-valued spin connection $\omega_-$: their BRST transformations define spinorial gauginos $\chi$ which are composites of the gravitino, its derivatives and other bosonic fields. In all cases, whenever $\phi$ is $\ig$-trivial, the topological gaugino must have the form~\eqref{topgauginostructure} in order to satisfy the consistency condition~\eqref{lambdaclosed}.

Invariant traces $\Tr \lambda_-^n$ are all $\ig$-closed.  For $n=2$ they might be non-trivial, since the Fierz identity~\eqref{zetafierz3} and the analogous one involving $\Gamma^{\mu\nu}$ ensure the existence of non-trivial $\ig$-classes at ghost number 2:
\begin{equations}
k^{(1)} &\equiv - \frac{1}{2} \big( \bar\zeta\, \Gamma_\mu \zeta \big) \d x^\mu \eqend{,} \\
k^{(2)} &\equiv \frac{1}{2} \big( \bar\zeta\, \Gamma_{\mu\nu} \zeta \big) \d x^\mu \d x^\nu \eqend{.}
\end{equations}
However there is no invariant $\ig$-cohomology at ghost number +3, and thus $\Tr \lambda_-^3$ is necessarily $\ig$-exact: it turns out that
\begin{equation}
\Tr \lambda^3_- = \ig \Tr \left[ \frac{\mathi}{4} \big( \bar\zeta\, \Gamma_\mu \chi \big) \big( \bar\chi\, \Gamma^\mu \Gamma_5 \chi \big) \sqrt{-g} \d^4 x \right] \eqend{.}
\end{equation}
Hence
\begin{equation}
\Omega^{(1)}_4 = \frac{\mathi}{4} \Tr \left[ \big( \bar\zeta\, \Gamma_\mu \chi \big) \big( \bar\chi\, \Gamma^\mu \Gamma_5 \chi \big) \sqrt{-g} \d^4 x \right] \eqend{,}
\end{equation}
and the supersymmetric anomaly cocycle is 
\begin{equation}
\mathcal{A}_5 = \Gamma_5(\mathbb{A}, \mathbb{F}) - \frac{\mathi}{4} \Tr \left[ \big( \bar\zeta\, \Gamma_\mu \chi \big) \big( \bar\chi\, \Gamma^\mu \Gamma_5 \chi \big) \sqrt{-g} \d^4 x \right] \eqend{.}
\label{superCSanomalybis}
\end{equation}

Let us summarize this discussion. A supergravity theory is characterized by a bilinear $\phi$ of ghost number 2. Its $S$ variation is always $\ig$-exact, but $\phi$ itself might or might not be $\ig$-exact. In all the examples we worked out explicitly, in any dimensions $\phi$ is $\ig$-exact for simple $\mathcal{N} = 1$ supergravities and non-trivial for extended supergravities --- although we do not know of an a priori argument for this to be so. In any case, for 4-dimensional $\mathcal{N} = 1$ new minimal supergravity $\phi$ is $\ig$-exact, and it can therefore be absorbed in a redefined YM connection~\eqref{Aminusconnection}. From the $S$ variation of the redefined connection, one then obtains a topological gaugino $\lambda_-$~\eqref{lambdaminusfromsaminus} which is $\ig$-closed. $\ig$-triviality of $\Tr \lambda_-^3$ in turn ensures that there exists a supersymmetrization of the familiar Chern--Simons anomaly functional~\eqref{superCSanomalybis} which is a BRST anomaly cocycle.

The supersymmetry anomaly cocycle is expressed in terms of the ``spinorial'' component $\chi$~\eqref{topgauginostructure} of the topological gaugino. When the YM field $A$ is the connection component of a super-YM multiplet coupled to supergravity, the spinorial gaugino is the usual gaugino, and the anomaly~\eqref{superCSanomalybis} is the well-known supersymmetric chiral anomaly. It describes the anomalies of  matter supersymmetric theories whose flavor symmetries are gauged by the classical external super-YM multiplet. In contrast, when the YM connection $A$ belongs to different supergravity multiplets, the ``spinorial gaugino'' is a composite field. In this case the BRST cocycle~\eqref{superCSanomalybis} is, apparently, something different than the standard chiral supersymmetric anomaly.
 
In the following Sections we will focus on $\mathcal{N} = 1$ new minimal supergravity~\cite{Ferrara:1988qxa}. It turns out that the spinorial gaugino $\chi$ for this model is proportional to the classical equations of motions for the gravitino. One might therefore suspect that for this theory the BRST cocycle~\eqref{superCSanomalybis} trivializes if one enlarges the space of fields of supergravity to include antifields, and that the anomaly is thus removable. We will verify that this is indeed the case: however this means that in order to remove such an anomaly, suitable deformations of the supersymmetry variations of the supergravity fields are required. Describing these deformations will be our goal.

\section{The BRST algebra in the \texorpdfstring{$B$}{B}-field sector}
\label{sec:Bsector}

The local bosonic symmetries of $\mathcal{N} = 1$ new minimal supergravity include --- beyond diffeomorphisms and YM symmetries --- also a vectorial gauge symmetry. This requires a slight generalization of the BRST equivariant formulation of supergravity that we outlined in Section~\ref{sec:sugrabrst}. Let 
\begin{equation}
V = V_\mu \d x^\mu
\end{equation}
be the anticommuting ghost field of ghost number 1 associated to the vector gauge symmetry, and
\begin{equation}
B = \frac{1}{2} B_{\mu\nu} \d x^\mu \d x^\nu
\end{equation}
the corresponding commuting gauge field of ghost number 0. Since the vectorial gauge symmetry is reducible, the BRST formulation requires also a scalar ghost-for-ghost field $q$ of ghost number 2. The BRST rules of the $B$ sector are
\begin{equations}[Brules]
s\, B &= - \mathcal{L}_\xi B - \d V - \bar\zeta\, \mathbf{\Gamma} \psi \eqend{,} \\
s\, V &= - \mathcal{L}_\xi V - \d q + \ig (B) - \frac{1}{2} \bar\zeta\, \mathbf{\Gamma} \zeta \eqend{,} \\
s\, q &= - \mathcal{L}_\xi q + \ig(V) \eqend{,}
\end{equations}
where we introduced the gravitino 1-form $\psi \equiv \psi_\mu \d x^\mu$ and defined the matrix-valued 1-form $\mathbf{\Gamma} \equiv \Gamma_\mu \d x^\mu$.

We have therefore two alternatives to extend the action of the equivariant BRST operator $S$ to the $B$ field. We could define $S$ equivariant with respect to \emph{all} local bosonic symmetries: diffeomorphisms, YM gauge transformations \emph{and} vector gauge transformations with ghosts $V^\mu$, resulting in
\begin{equation}
S\, B = - \bar\zeta \, \mathbf{\Gamma} \psi \eqend{.}
\end{equation}
If we denote by $\delta'_V$ the vector gauge transformation with parameter $V$, the equivariant BRST algebra relation
\begin{equation}
S^2 = \mathcal{L}_\gamma + \delta^\text{\tiny YM}_{\ig(A) + \phi} + \delta'_{\ig(B) + k^{(1)}}
\label{BRSTequivariantSone}
\end{equation}
holds with this choice on all fields, except the ghosts associated to bosonic gauge symmetries, i.e. $\xi$, $c$, $V$ and $q$. Beyond the conditions~\eqref{gammaconsistency} and~\eqref{phiconsistency}, consistency of~\eqref{BRSTequivariantSone} requires also
\begin{equation}
S\, \big( \ig(B) + k^{(1)} \big) = 0 \eqend{.}
\label{Bconsistency}
\end{equation}
Alternatively, we could as well define a BRST operator $\tilde{S}$ equivariant only with respect to diffeomorphisms and YM symmetries:
\begin{equations}
\tilde{S}\, B &=  - \d V - \bar\zeta \, \mathbf{\Gamma} \psi \eqend{,} \\
\tilde{S}\, V &= - \d q + \ig(B) - \frac{1}{2} \bar\zeta \, \mathbf{\Gamma} \zeta \eqend{,}
\end{equations}
in which case the BRST algebra would be
\begin{equation}
\tilde{S}^2 = \mathcal{L}_\gamma + \delta^\text{\tiny YM}_{\ig(A) + \phi} \eqend{,}
\end{equation}
which would hold on all fields --- \emph{including} $B$ and $V$ ---  with the exception of the ghosts $\xi$ and $c$. In the following, we will find more convenient to use the fully equivariant $S$.

The BRST action in the $B$ sector becomes more transparent if we collect the various fields $q$, $V$ and $B$ in a single generalized form of total fermionic number 2:
\begin{equation}
\mathbb{B} = B + V + q \eqend{.}
\end{equation}
Then the BRST variations~\eqref{Brules} are equivalent to
\begin{equation}
\delta \, \mathbb{B} + \frac{1}{2} \mathbf{\bar\Psi} \, \mathbf{\Gamma} \, \mathbf{\Psi} = H \eqend{,}
\label{Bsuperrules}
\end{equation}
where
\begin{equation}
H \equiv \d B + \frac{1}{2} \bar\psi\, \mathbf{\Gamma}\, \psi \eqend{,}
\label{Hdef}
\end{equation}
and we have introduced, in a way analogous to Eq.~\eqref{superconnection}, the generalized commuting super-gravitino form of ghost number 1
\begin{equation}
\mathbf{\Psi} \equiv \zeta + \psi \eqend{.}
\end{equation}
The 3-form $H$ transforms nicely under supersymmetry. To see this, it is instructive to start from the BRST properties of the vierbein. Let us introduce the gauge covariant coboundary operator
\begin{equation}
\hat{\delta} \equiv s + \mathcal{L}_\xi + D - \ig \eqend{,}
\end{equation}
where $D = \d + [A,\cdot]$ is the exterior derivative covariant with respect to local YM symmetries. We have
\begin{equation}
\hat{\delta} e^a =  - \mathbf{\bar\Psi}\, \Gamma^a\, \mathbf{\Psi} \eqend{.}
\label{supertorsion4}
\end{equation}
The covariant coboundary operator squares to a YM gauge transformation:
\begin{equation}
\hat{\delta}^2 = \delta^\text{\tiny YM}_{\mathbb{F}} \eqend{.}
\end{equation}
Hence
\begin{equation}
\hat{\delta}^2 e^a = \mathbb{R}^a{}_b \, e^b \eqend{,}
\end{equation}
where $\mathbb{R}^{ab}$ is the Lorentz component of the super-curvature $\mathbb{F}$. Therefore we obtain
\begin{splitequation}
\mathbb{R}_{ab}\, e^a\, e^b &= - e_a \, \hat{\delta}\, \mathbf{\bar\Psi}\, \Gamma^a\, \mathbf{\Psi} = \delta\,\bigl( e_a\, \mathbf{\bar\Psi} \,  \Gamma^a\, \mathbf{\Psi}\bigr) - \bigl( \mathbf{\bar\Psi} \,  \Gamma_a\, \mathbf{\Psi} \bigr) \bigl( \mathbf{\bar\Psi} \,  \Gamma^a\, \mathbf{\Psi} \bigr) \\
&= \delta\,\bigl( e_a\, \mathbf{\bar\Psi} \, \Gamma^a\, \mathbf{\Psi} \bigr) \eqend{,}
\label{supertorsionone}
\end{splitequation}
where we made use of the super-Fierz identity which generalizes~\eqref{zetafierz3}:
\begin{equation}
\bigl( \mathbf{\bar\Psi} \,  \Gamma_a\, \mathbf{\Psi} \bigr) \bigl( \mathbf{\bar\Psi} \,  \Gamma^a\, \mathbf{\Psi} \bigr) = 0 \eqend{.}
\label{superfierzone}
\end{equation}
Eq.~\eqref{supertorsionone} can be considered as the analog of the first Bianchi identity for the generalized curvature: the ghost number 0 component of this equation it is just the familiar relation connecting the derivative of the torsion to the cyclic sum of the curvature tensor components. In this sense
\begin{equation}
\mathbb{T} \equiv \frac{1}{2} \mathbf{\bar\Psi}\, \mathbf{\Gamma}\, \mathbf{\Psi}
\label{generalizedtorsion}
\end{equation}
is the generalized torsion, and the first super-Bianchi identity reads
\begin{equation}
\frac{1}{2} \mathbb{R}_{ab}\, e^a\, e^b  = \delta\, \mathbb{T} \eqend{.}
\label{supertorsiononebis}
\end{equation}
The BRST rules for $B$~\eqref{Bsuperrules} imply that the generalized torsion is cohomologous to a 3-form of ghost number 0:
\begin{equation}
\delta\, \mathbb{T} = \delta H \eqend{,}
\end{equation}
so that the first super-Bianchi identity takes the final form
\begin{equation}
\delta H = \frac{1}{2} \mathbb{R}_{ab}\, e^a\, e^b \eqend{.}
\label{supertorsiontwo}
\end{equation}

\section{The supersymmetry BRST anomaly cocycle for\texorpdfstring{\\}{} new minimal \texorpdfstring{$\mathcal{N} = 1$}{N=1} supergravity}
\label{sec:newN1BRSTrules}

Let us now summarize the equivariant BRST rules of the new minimal $\mathcal{N} = 1$ supergravity multiplet~\cite{Ferrara:1988qxa}:
\begin{equations}[summaryN1newmin]
S\, \zeta &= \ig(\psi) \eqend{,} \\
S\, e^a &= - 2 \bar\zeta \, \Gamma^a \psi \eqend{,} \\
S\, \psi &= - D^+ \zeta \eqend{,} \\
S\, B &= - \bar\zeta \, \mathbf{\Gamma} \psi \eqend{,} \\
S \, A^- &= \mathi \, \bar\zeta \, \Gamma_5 \, \mathbf{\Gamma} \, \Gamma^{\mu\nu} \tilde{\psi}_{\mu\nu} \eqend{,} \\
S\, \omega_-^{ab} &= 2 \bar\zeta \, \mathbf{\Gamma} \, \tilde\psi^{ab} \eqend{,} \\
S\, H &= \bar\zeta \, \mathbf{\Gamma} \, \tilde\psi \eqend{,}
\end{equations}
where the Lorentz-covariant derivative $D$ acts on spinors according to
\begin{equation}
D\, \zeta \equiv \d \zeta - \frac{1}{4} \omega^{ab}\, \Gamma_{ab}\,\zeta - \frac{\mathi}{2} A \,\Gamma_5\, \zeta \eqend{,}
\label{deflorentzd}
\end{equation}
and the following combinations are useful:
\begin{equations}[summaryN1newmincombinations]
H &\equiv \d B - \frac{1}{2} \bar\psi \, \mathbf{\Gamma} \, \psi \eqend{,} \quad H^\lambda = \frac{1}{6} \epsilon^{\lambda\mu\nu\rho} H_{\mu\nu\rho} \eqend{,} \\
A_\mu^- &\equiv A_\mu - 3 H_\mu \eqend{,} \hspace{2.5em} A_\mu^+ \equiv A_\mu - H_\mu \eqend{,} \\
\omega^\pm_\mu{}^{ab} &= \omega_\mu{}^{ab} \pm H_\mu{}^{ab} \eqend{,} \\
\tilde\psi &\equiv D^+\psi \eqend{,} \hspace{4.5em} \tilde\psi^{ab} \equiv e^{a\mu} e^{b\nu} \tilde{\psi}_{\mu\nu} \eqend{.}
\end{equations}
The Lorentz-covariant derivatives $D^\pm$ are defined in analogy to~\eqref{deflorentzd}, but using the spin connection $\omega^\pm_\mu{}^{ab}$ and the gauge field $A^\pm_\mu$.

Comparing Eqs.~\eqref{summaryN1newmin} with Eq.~\eqref{topgauginostructure}, we conclude that the components of the ``spinorial'' composite gaugino associated to the $\UR$ and the Lorentz gauge algebras are, respectively
\begin{equation}
\chi^{\scriptscriptstyle \UR} = 2\,\Gamma^{\mu\nu} \tilde{\psi}_{\mu\nu} \quad \text{and}\quad \chi^{ab} = 2 \tilde\psi^{ab} \eqend{.}
\end{equation}

The Lorentz components of the spinorial composite gaugino do not contribute to the supersymmetric anomaly cocycle~\eqref{superCSanomalybis} for group theoretical reasons: the completely symmetric primitive invariant (symmetric 3-index symbol) $d_{abc}$ vanishes for $\mathrm{SO}(1,3)$. However, the axial $\UR$ component gives a non-vanishing contribution to the supersymmetric anomaly cocycle:
\begin{equation}
\label{anomalyN=1explicitform}
\mathcal{A} = c \, ( F^- )^2 + 2 A^- \lambda^- F^- + \frac{\mathi}{24} \lambda^- \left( \bar\chi \, \Gamma^\nu \, \Gamma_5 \, \chi \right) \epsilon_{\nu\rho\sigma\lambda} \sqrt{-g} \d x^\rho \d x^\sigma \d x^\lambda
\end{equation}
where we have identified $\chi$ and $\lambda^-$ with their $\UR$ components in order to simplify the notation.

The BRST operator~\eqref{summaryN1newmin} of new minimal $\mathcal{N} = 1$ supergravity is nilpotent on the space of fields, without the need to introduce antifields, and therefore the BRST cohomology problem on the space of fields is well defined. The anomaly cocycle associated to the $\UR$ axial gauge field is a non-trivial element of this cohomology modulo the exterior differential:
\begin{equation}
s\, \mathcal{A} = - \d \big[ i_\xi\, \mathcal{A} + 2\, c \, F^- \lambda^- + A^- (\lambda^-)^2 \big] \eqend{.}
\label{Sanomalyinvariance}
\end{equation}
However, as we will explain in the following Sections, it becomes trivial when we enlarge the field space to include antifields.

\section{``Evanescent'' Anomalies}
\label{sec:evanescent anomalies}

Consider a supersymmetric ``matter'' quantum field theory whose currents are coupled to the (classical) fields of supergravity. Let $\phi^i$ denote the collections of supergravity fields, and let $\Gamma_\text{eff}[\phi]$ be the effective action of the matter theory, i.e., the generating functional of correlation functions of the currents that are coupled to the $\phi^i$. In this Section, we denote by $s_0$ the full BRST operator of $\mathcal{N} = 1$ new minimal supergravity, which in the previous Sections was denoted by $s$. $s_0$ is nilpotent on the space of supergravity and ghost fields. We will call this space the ``small''  field space, to distinguish it from the ``big'' space involving both fields and antifields. We assume that $\Gamma_\text{eff}[\phi]$ is anomalous:
\begin{equation}
s_0 \, \Gamma_\text{eff}[\phi] = t \int\! \mathcal{A}[\phi] \eqend{,}
\label{AnomalyAnti}
\end{equation}
where $\mathcal{A}[\phi]$ is a non-trivial element of the $s_0$ cohomology (modulo the exterior differential) on the ``small'' field space. We introduced a formal parameter $t$, which in the case at hand is $\mathcal{O}(\hbar)$ since the matter SQFT is assumed to be classically supersymmetric. It should be kept in mind that we also take $\Gamma_\text{eff}[\phi]$ to be of order $\mathcal{O}(t)$.
 
Let us also assume that $\mathcal{A}[\phi]$ vanishes on the subspace of the equations of motion associated to some \emph{local}, $s_0$-invariant action $\Gamma_{0}[\phi]$ of the supergravity fields $\phi^i$. In this situation we will say that the anomaly $\mathcal{A}[\phi]$ is ``evanescent''. We further assume that $\Gamma_0[\phi]$ does not depend on the ghost fields, or in other words that $\Gamma_0[\phi]$ can be identified with (any) classical supergravity action. Being both $s_0$-invariant and ghost-independent, $\Gamma_{0}[\phi]$ is also invariant under the bosonic gauge symmetries encoded in $s_0$ --- in the context of $\mathcal{N} = 1$ new minimal supergravity this means that $\Gamma_{0}[\phi]$ is invariant under diffeomorphisms, local YM symmetries and vectorial symmetries associated to $B$. We can therefore write
\begin{equation}
\int\! \mathcal{A}[\phi] = - t\, S_1\, \Gamma_0[\phi] \eqend{,} \qquad s_0\, \Gamma_0[\phi] = 0 \eqend{,}
\label{Anomalyevanescent}
\end{equation}
where $S_1$ is an odd ghost number 1 operator acting locally on $\phi^i$, and $\Gamma_0[\phi]$ is $\mathcal{O}(t^0)$. Since $\Gamma_0[\phi]$ is gauge invariant, $S_1$ is gauge covariant:
\begin{equation}
\{ S_1, \mathcal{L}_\xi \} = \mathcal{L}_{S_1 \xi} \eqend{,} \qquad \{ S_1, \delta^\text{\tiny YM}_c \} = \delta^\text{\tiny YM}_{S_1\,c} \eqend{,} \qquad \{ S_1, \delta'_V \} = \delta'_{S_1 V} \eqend{.}
\end{equation}
Let us observe that since $\Gamma_0[\phi]$ is ghost independent, the action of $S_1$ on the ghost is not determined by Eq.~\eqref{Anomalyevanescent}: it will have to be fixed by consistency. Putting Eqs.~\eqref{Anomalyevanescent} and~\eqref{AnomalyAnti} together, we obtain
\begin{equation}
s_0 \, \Gamma_\text{eff}[\phi] = - t\, S_1\, \Gamma_0[\phi] \eqend{.}
\label{s1anomaly}
\end{equation}
Because of the nilpotency of $s_0$ and $s_0$-invariance of $\Gamma_0[\phi]$, we also have
\begin{equation}
\{ s_0, S_1 \} \, \Gamma_0[\phi] = 0 \eqend{.}
\label{s0s1consistency}
\end{equation}
Hence $s_0$ and $S_1$ must anticommute --- up to bosonic gauge symmetries of the action $\Gamma_{0}[\phi]$. If we introduce the equivariant $S_0$ by
\begin{equation}
s_0 = - \mathcal{L}_\xi - \delta^\text{\tiny YM}_c - \delta'_V + S_0 \eqend{,}
\end{equation}
then the consistency equation~\eqref{s0s1consistency} reads
\begin{equation}
\{ S_0, S_1 \} = \mathcal{L}_{\gamma_1} + \delta^\text{\tiny YM}_{\phi_1} + \delta'_{k_1} \eqend{,}
\label{s0s1consistencybis}
\end{equation}
where $\gamma_1$, $\phi_1$ and $k_1$ are, respectively, a ghost number 2 vector, a YM Lie algebra-valued scalar, and a one-form. Let us now introduce the deformed equivariant BRST operator, depending on the formal parameter $t$:
\begin{equation}
S_t \equiv S_0 + t S_1 + t^2 S_2 + \ldots \eqend{.}
\label{stdeformed}
\end{equation}
The consistency equation~\eqref{s0s1consistencybis} reads, up to $\mathcal{O}(t^2)$ 
\begin{equation}
S_t^2 = \mathcal{L}_{\gamma_t} + \delta^\text{\tiny YM}_{i_{\gamma_t}(A) + \phi_t} + \delta'_{i_{\gamma_t}(B) + k_t} \eqend{,}
\label{Stdeformedalgebra}
\end{equation}
where
\begin{equations}
\gamma_t &\equiv \gamma + t \gamma_1 + \mathcal{O}(t^2) \eqend{,} \\
\phi_t &\equiv \phi + t \phi_1 + \mathcal{O}(t^2) \eqend{,} \\
k_t &\equiv k + t k_1 + \mathcal{O}(t^2) \eqend{.}
\end{equations}
Moreover, the total effective action
\begin{equation}
\Gamma[\phi] = \Gamma_0[\phi] + \Gamma_\text{eff}[\phi] + \mathcal{O}(t^2)
\label{effectivedeformed}
\end{equation}
differs from the original one by a local term $\Gamma_0[\phi]$ and is $S_t$ invariant up to this order:
\begin{equation}
S_t \, \Gamma[\phi] = \mathcal{O}(t^2) \eqend{.}
\end{equation}
If we can continue this procedure to all orders in $t$, by adding terms $S_n$ to $S_t$ of higher order in $t$, and \emph{local} terms of higher order in $t$ to the effective action
\begin{equation}
\Gamma[\phi] = \Gamma_0[\phi] + \Gamma_\text{eff}[\phi] + t^2 \Gamma_2[\phi] + \ldots \eqend{,}
\end{equation}
we have then removed the original --- ``evanescent'' --- anomaly. We end up with an effective action $\Gamma[\phi]$, which differs from the original one only by the addition of local terms and which is invariant, that is non-anomalous, under the new \emph{deformed} supergravity BRST operator $S_t$.

Let us discuss what kind of restrictions nilpotency and other physical requirements put on the deformed algebra~\eqref{Stdeformedalgebra}. As we discussed in Section~\ref{sec:sugrabrst}, nilpotency of the BRST operator requires that
\begin{equation}
S_t \, \gamma_t = 0 \eqend{,} \qquad S_t\, \left[ \phi_t + i_{\gamma_t}(A) \right] = 0 \eqend{,} \qquad S_t\, \left[ k_t + i_{\gamma_t}(B) \right] = 0 \eqend{.}
\end{equation}
We also established that the BRST rules for the supersymmetry ghost and the vierbein are universal, i.e., they should be valid for any supergravity theory. Hence we should require that the BRST rules for $\zeta$ and $e^a$ be unchanged by the deformation:
\begin{equation}
S_t \, \zeta = S_0\, \zeta = \ig(\psi) \eqend{,} \qquad S_t\, e^a = S_0\, e^a = - 2 \bar\zeta\, \Gamma^a \psi \eqend{.}
\label{Stconstraints}
\end{equation}
Consequently the ghost number 2 vector bilinear that appears in the algebra of the deformed supergravity BRST operator $S_t$ should also be unaffected by the deformation:
\begin{equation}
\gamma^\mu_t = \gamma^\mu \eqend{.}
\end{equation}
The other 2 bilinears, $\phi_t$ and $k_t$, are instead not universal: therefore they might be deformed, subject to the constraints~\eqref{Stconstraints}. In conclusion, the deformed algebra of the equivariant BRST operator will read
\begin{equation}
S^2_t = \mathcal{L}_\gamma + \delta^\text{\tiny YM}_{\ig (A) + \phi_t} + \delta'_{\ig (B) + k_t} \eqend{.}
\label{Stdeformedalgebrabis}
\end{equation}
At first order in $t$, we have seen that this is equivalent to the existence of an operator $S_1$ of ghost number 1, which anticommutes with $S_0$ up to gauge transformations:
\begin{equation}
\{ S_1, S_0 \} = \delta^\text{\tiny YM}_{\phi_1} + \delta'_{k_1} \eqend{.}
\label{s0s1consistencytris}
\end{equation}
On the other hand, if $S_1$ were given by a $S_0$-commutator
\begin{equation}
S_1 = \{ S_0, L_1\}
\end{equation}
for some local operator $L_1$ of ghost number 0, then we would obtain
\begin{equation}
0 = S_0 \, \Gamma_\text{eff}[\phi] + t S_1\, \Gamma_0[\phi] = S_0\, \bigl( \Gamma_\text{eff}[\phi] + t L_1\, \Gamma_0[\phi] \bigr) \eqend{.}
\end{equation}
In other words, $S_1$ operators which are given by $S_0$-commutators trigger trivial deformations of the original $S_0$: they correspond to anomalies which are trivial in the ``small'' field space and to $S_0$-invariant effective actions differing from the original one only by local terms. Hence, if the anomaly is a non-trivial element of the $S_0$ cohomology in the ``small'  field space, then $S_1$ is certainly not a $S_0$-commutator.
 
It is therefore useful to introduce the notion of $S_0$-cohomology on the space of \emph{operators} acting as \emph{local derivatives} on field space. The action of $S_0$ on local functional derivatives is given by the commutator (respectively, anticommutator) for even (respectively, odd) derivatives. Closed operators are operators which (anti-)commute with $S_0$ up to gauge transformations, and trivial ones are those which are $S_0$-(anti-)commutators up to gauge transformations. We will refer to this cohomology as the $S_0$-operatorial cohomology. We have just seen that consistent, non-trivial deformations $S_t$ require the existence of a $S_1$ which is a non-trivial element of ghost number +1 of this operatorial $S_0$-cohomology.

At higher order in $t$, the consistency condition~\eqref{Stdeformedalgebrabis} imposes restrictions on the operators $S_n$ with $n > 1$. For example, at order 2 in $t$, Eq.~\eqref{Stdeformedalgebrabis} gives
\begin{equation}
S_1^2 + \{ S_0, S_2 \} = \delta^\text{\tiny YM}_{\phi_2} + \delta'_{k_2} \eqend{.}
\label{St2consistency} 
\end{equation}
This equation says that the ghost number 2 operator $S_1^2$ is a trivial element of the $S_0$ operatorial cohomology at ghost number 2. On the other hand, the first-order condition~\eqref{s0s1consistencybis} ensures that the commutator $[S_0, S_1^2]$ vanishes up to gauge transformations:
\begin{equation}
\left[ S_0, S_1^2 \right] = - \delta^\text{\tiny YM}_{S_1\, \phi_1} - \delta'_{S_1\, k_1} \eqend{.}
\end{equation}
In other words, if $S_1$ exists, then its square $S_1^2$ is $S_0$-closed. A consistent deformation $S_t$ at the next order requires that $S_1^2$ be a trivial element of the $S_0$ operatorial cohomology at ghost number +2. One can check that all the higher-order conditions are analogous statements on $S_0$ operatorial cohomologies at higher ghost numbers. For example, the consistency condition at third order in $t$ amounts to the requirement that the $S_0$-closed operator $\{S_1, S_2\}$ of ghost number 3 be $S_0$-trivial.
 
Summing up, a sufficient condition for the existence of a consistent deformation $S_t$ that removes the original ``evanescent'' anomaly is the validity of the following two facts:
\begin{enumerate}
\item[a)] the existence of a non-trivial element of the $S_0$-operatorial cohomology at ghost number 1, and
\item[b)] the emptiness of the same cohomology at all higher ghost numbers.
\end{enumerate}

\section{The deformation}
\label{sec:deformation}

We are now finally in the position to determine the first-order deformation of the equivariant BRST operator of $\mathcal{N} = 1$ new minimal supergravity that removes the original ``evanescent'' anomaly. The supergravity Lagrangian density invariant under $S_0$ is~\cite{Ferrara:1988qxa}
\begin{equation}
\mathcal{L} = \abs{e} \bigl[ R(\omega) - 4 \bar\psi_\mu \Gamma^{\mu\nu\rho}\, D_\nu(\omega) \,\psi_\rho - 6 H^\mu H_\mu + 4 A_\mu H^\mu \bigr] \eqend{,}
\label{sugralagrangian}
\end{equation}
where we recall that
\begin{equation}
D(\omega) = \d x^\nu D_\nu (\omega) = \d x^\nu \left( \partial_\nu + \frac{1}{4} \omega_\nu{}^{ab} \Gamma_{ab} \right)
\end{equation}
is the Lorentz-covariant derivative and
\begin{equation}
H^\mu = \frac{1}{2 \abs{e}} \epsilon^{\mu\nu\rho\sigma} \bigl( \partial_\nu B_{\rho\sigma} + \bar\psi_\nu \Gamma_\rho\, \psi_\sigma \bigr)
\end{equation}
is the vector dual of the 3-form $H$. The spin connection $\omega_\mu{}^{ab} = \omega_\mu{}^{ab}(e, \psi)$ is determined by its equation of motion that follows from the Lagrangian~\eqref{sugralagrangian}, and given by
\begin{equation}
D(\omega) e^a + \frac{1}{2} \bar\psi\, \Gamma^a \psi = 0 \eqend{,}
\label{spinconnectioneom}
\end{equation}
which is the 2-form component of the super-torsion constraint~\eqref{supertorsion4}.

Taking $A_\mu$, $B_{\mu\nu}$, $e^a_\mu$ and $\psi_\mu$ as independent fields, the equations of motion other than the Einstein equations are
\begin{equations}
\frac{1}{\abs{e}} \frac{\delta \mathcal{L}}{\delta A_\mu} &= 4 H^\mu \eqend{,} \\
\frac{1}{\abs{e}} \frac{\delta \mathcal{L}}{\delta B_{\mu\nu}} &= \epsilon^{\mu\nu\rho\sigma} F^-_{\rho\sigma} \eqend{,} \\
\frac{1}{\abs{e}} \frac{\delta \mathcal{L}}{\delta \bar\psi_\mu} &= - 8 \big[ \Gamma^{\mu\nu\rho} D^+_\nu \psi_\rho + \frac{1}{2} H_\nu \, \epsilon^{\mu\nu\rho\sigma} \Gamma_\sigma \psi_\rho + \mathi H^\mu \, \Gamma^\rho \, \Gamma_5 \psi_\rho - \mathi H^\rho \, \Gamma^\mu \, \Gamma_5 \psi_\rho \big] \eqend{,}
\label{eomssugramin}
\end{equations}
where we recall that the Lorentz-covariant derivative $D^+$ is defined in analogy to~\eqref{deflorentzd}, but using the spin connection $\omega^\pm_\mu{}^{ab}$ and the gauge field $A^\pm_\mu$ defined in Eqs.~\eqref{summaryN1newmincombinations}.

The generalized form
\begin{equation}
\mathcal{A}_5 = \Gamma_5(\mathbb{A}, \mathbb{F}) + \frac{\mathi}{24} \lambda^- \left( \bar\chi \, \Gamma^\nu \, \Gamma_5 \, \chi \right) \epsilon_{\nu\rho\sigma\lambda} \sqrt{-g} \d x^\rho \d x^\sigma \d x^\lambda
\end{equation}
describes the $\UR$ component of the anomaly~\eqref{superCSanomalybis} of $\mathcal{N}=1$ new minimal supergravity. In this formula, $\lambda_- = \lambda^-_\mu \d x^\mu $ is the $\UR$ component of the topological gaugino\footnote{In this section $A_-$, $\lambda_-$, $F_-$ and $\phi$ refer to the $\UR$ components of the YM topological multiplet. The corresponding Lorentz components will be denoted by $\omega_-^{ab}$, $\lambda_-^{ab}$, $R_-^{ab}$  and $\phi^{ab}$. Of course at zeroth order in $t$ we have $\phi = \phi^{ab} = 0$.}
\begin{equation}
\lambda^-_\mu = \bar\zeta\, \Gamma_\mu \chi \eqend{,}
\end{equation}
and $\chi$ is the spinorial gaugino
\begin{splitequation}
\chi &= - 2 \mathi \, \Gamma_5\, \Gamma^{\rho\sigma}\, D^+_\rho \psi_\sigma = \frac{\mathi}{8} \Gamma_5\, \Gamma_\rho \frac{\delta \mathcal{L}}{\delta \bar\psi_\rho} + 3 H^\rho \, \psi_\rho \\
&= \frac{\mathi}{8} \Gamma_5\, \Gamma_\rho \frac{\delta \mathcal{L}}{\delta \bar\psi_\rho} + \frac{3}{4} \frac{\delta \mathcal{L}}{\delta A_\mu} \psi_\mu \eqend{,}
\end{splitequation}
which is proportional to the equations of motion for $\psi_\mu$ and $A_\mu$.

The 4-form component of the anomaly polyform $\mathcal{A}_5$ gives the anomaly density~\eqref{anomalyN=1explicitform}
\begin{equation}
\mathcal{A} = \frac{\abs{e}}{4} \left[ \bigl( c\, F^-_{\mu\nu} - 2 A^-_{[\mu}\,\lambda^-_{\nu]} \bigr) F^-_{\rho\sigma} \, \epsilon^{\mu\nu\rho\sigma} - 2 A^-_{\mu} \lambda^-_{\nu} F^-_{\rho\sigma} \, \epsilon^{\mu\nu\rho\sigma} + 3 \mathi \, \left( \bar\chi\, \Gamma^\mu\, \Gamma_5 \chi \right) \lambda^-_\mu \right] \eqend{.}
\label{A4anomalyN1}
\end{equation}
Comparing with Eq.~\eqref{eomssugramin}, it is clear that this anomaly is ``evanescent'': it vanishes when the equations of motion of $\psi_\mu$, $B_{\mu\nu}$ and $A_\mu$ are satisfied.  We can schematically write\footnote{To avoid cluttering, we drop all the space-time indices, including the integration.}
\begin{equation}
\mathcal{A} = a_1^i[\phi] \frac{\delta \Gamma_0[\phi]}{\delta \phi^i} + a_2^{ij}[\phi] \frac{\delta \Gamma_0[\phi]}{\delta \phi^i} \frac{\delta \Gamma_0[\phi]}{\delta \phi^j} + a_3^{ijk}[\phi] \frac{\delta \Gamma_0[\phi]}{\delta \phi^i} \frac{\delta \Gamma_0[\phi]}{\delta \phi^j} \frac{\delta \Gamma_0[\phi]}{\delta \phi^k} \eqend{,}
\label{anomalyambiguityone}
\end{equation}
where no terms of higher order than cubic appear. There is a certain degree of ambiguity in reading off from this formula the action of the deformation $S_1\, \phi^i $: let us briefly pause to discuss it.

Let us first observe that we can \emph{define} the functionals $a_2^{ij}[\phi]$ and $a_3^{ijk}[\phi]$ in Eq.~\eqref{anomalyambiguityone} to be either symmetric or antisymmetric with respect to the exchange of \emph{any} pair $(i,j)$ of two indices: antisymmetric if $i$ and $j$ both correspond to fermionic fields, symmetric in the other cases. We could therefore take $S_1$ to be 
\begin{equation}
- S_1\, \phi^i = a_1^i[\phi] + a_2^{ij}[\phi] \frac{\delta \Gamma_0[\phi]}{\delta \phi^j} + a_3^{ijk}[\phi] \frac{\delta \Gamma_0[\phi]}{\delta \phi^j} \frac{\delta \Gamma_0[\phi]}{\delta \phi^k}
\label{S1symmetric}
\end{equation}
with $a_2^{ij}[\phi]$ and $a_3^{ijk}[\phi]$ completely (anti-)symmetric under exchange of field indices. Such a $S_1$ would certainly satisfy the defining equation~\eqref{s1anomaly}.

However, $S_1$ is only defined up to invariances of $\Gamma_0[\phi]$. These include of course local symmetries of $\Gamma_0$: diffeomorphisms and YM gauge symmetries. But if we allow in $S_1$ for terms bi- and trilinear in the equations of motion as in Eq.~\eqref{anomalyambiguityone}, then we can add to $S_1$ ``trivial'' symmetries of $\Gamma_0$ of the form:\footnote{``Trivial'' symmetries of this kind are sometimes called zilch symmetries. Of course this kind of triviality has nothing to do with cohomological BRST triviality.}
\begin{equation}
S^\text{trivial}_1\, \phi^i = b_2^{ij}[\phi] \frac{\delta \Gamma_0[\phi]}{\delta \phi^j} + b_3^{ijk}[\phi] \frac{\delta \Gamma_0[\phi]}{\delta \phi^j} \frac{\delta \Gamma_0[\phi]}{\delta \phi^k} \eqend{,}
\label{S1ambiguity}
\end{equation}
where $b_2^{ij}[\phi]$ and $b_3^{ijk}[\phi]$ are functionals with the \emph{wrong} kind of symmetry under the exchange of any pair of two indices $i$ and $j$, i.e., with $b_2^{ij}[\phi]$ and $b_3^{ijk}[\phi]$ symmetric under the exchange of any two indices $i$ and $j$ corresponding to fermionic fields, and antisymmetric in all other cases. For example, any $b_2^{ij}$ satisfying
\begin{equation}
b_2^{ij}[\phi] = - (-1)^{n_i n_j} \, b_2^{ji}[\phi]
\label{wrongsymmetry}
\end{equation}
where $n_i = 0$ ($n_i = 1$) for bosonic (fermionic) fields, corresponds to a ``trivial'' symmetry, and analogously for $b_3^{ijk}[\phi]$.

The anticommutator of $S_0$ with any such ``trivial'' contribution to $S_1$ is also a ``trivial'' symmetry of $\Gamma_0$. We have seen that the removal of the anomaly requires that the anticommutator of $S_0$ with $S_1$ only contains genuine symmetries of the action $\Gamma_0$ --- that is, gauge symmetries and diffeomorphisms. One can therefore expect that, for this to be the case, $S_1$ should not contain any  ``trivial'' terms of the type~\eqref{S1ambiguity}. This is indeed the case: we have verified explicitly that the ambiguity in the definition of $S_1$ is completely fixed by requiring the anticommutator of $S_0$ and $S_1$ to satisfy Eq.~\eqref{s0s1consistencybis}, and the resulting  $S_1$ has the form~\eqref{S1symmetric} with $a^{ij}_2$ and $a^{ijk}_3$ completely (anti-)symmetric.

In conclusion, we can write
\begin{equation}
\int \mathcal{A} = - \int \left( S_1\, B_{\mu\nu} \right) \frac{\delta \Gamma_0}{\delta B_{\mu\nu}} + \left( S_1\, \psi_\mu \right) \frac{\delta \Gamma_0}{\delta \psi_\mu} + \left( S_1\, A_\mu \right) \frac{\delta \Gamma_0}{\delta A_\mu}
\label{anomalyfromdeformation}
\end{equation}
with
\begin{equations}[FinalS1]
S_1\, B_{\mu\nu} &= - \frac{1}{4} \bigl[ c\, F^-_{\mu\nu} - 2 A^-_{[\mu} \lambda^-_{\nu]} \bigr] \eqend{,} \\
S_1\, \psi_\mu &= - \frac{\mathi}{2} \Theta_\alpha \Gamma_\mu\, \Gamma^\alpha\, \Gamma_5\, \zeta + \frac{1}{16} \left( \bar\chi \chi \right) \Gamma_\mu\, \zeta - \frac{1}{16} \left( \bar\chi\, \Gamma_5 \chi \right) \Gamma_\mu\, \Gamma_5\, \zeta \eqend{,} \\
S_1\, A_\mu &= 3 \Theta_\alpha \bar\zeta\, \Gamma^\alpha \psi_\mu + \frac{3}{8} \, \mathi \bigl[ \left( \bar\chi \chi \right) \bar\zeta\, \Gamma_5 \psi_\mu - \left( \bar\chi\, \Gamma_5 \chi \right) \bar\zeta\,\psi_\mu\bigr] \eqend{,}
\end{equations}
where
\begin{equation}
\Theta \equiv \Theta_\alpha \d x^\alpha = - \frac{1}{8} \left[ A^-_\nu F^-_{\rho\sigma} \, \epsilon_{\alpha}{}^{\nu\rho\sigma} + \mathi \, \left( \bar\chi\, \Gamma_\alpha\, \Gamma_5 \chi \right) \right] \d x^\alpha \eqend{.}
\label{Thetavectordef}
\end{equation}
This deformation $S_1$ satisfies the integrability condition~\eqref{Stdeformedalgebrabis} resp.~\eqref{s0s1consistencytris}: the first-order deformation of the YM bilinear $\phi_1$ is non-trivial along the $\UR$ and the Lorentz directions, and the corresponding deformations $\phi^{(R)}_1$ and $\phi^{ab}_1$ read
\begin{equations}[Finalphi1deformation]
\phi^{(R)}_1 &= 3 \ig(\Theta) \eqend{,} \\
\phi^{ab}_1 &= \ig\left( \epsilon^{ab}{}_{\mu\nu} \, \Theta^\nu \d x^\mu \right) - \frac{1}{8} \left( \bar\chi \chi \right) \bar\zeta \, \Gamma^{ab} \zeta + \frac{1}{8} \left( \bar\chi\, \Gamma_5 \chi \right) \bar\zeta \, \Gamma^{ab}\, \Gamma_5 \zeta \eqend{.}
\end{equations}
The deformation of the bilinear associated to the gauge transformations of the antisymmetric $B$ field is instead
\begin{equation}
k_1 = - \frac{1}{4} c \, \lambda^- \eqend{.}
\label{Finalk1deformation}
\end{equation}

\section{The torsion constraint}
\label{sec:renormalizedtorsion}
 
The generalized first Bianchi identity~\eqref{supertorsionone}, \eqref{supertorsiononebis}
\begin{equation}
\mathbb{R}^{ab}\, e_a\, e_b = \delta\,\bigl( \mathbf{\bar\Psi} \, \mathbf{\Gamma}\, \mathbf{\Psi} \bigr) = \delta\, \mathbb{T}
\end{equation}
relies on the ``universal''  BRST rules, those for $\zeta$ and $e^a$. Therefore this equation must hold also for the deformed $S_t$.

At zeroth order in $t$, we also have the equation
\begin{equation}
\frac{1}{2} \mathbf{\bar\Psi} \, \mathbf{\Gamma}\, \mathbf{\Psi} =  H - \delta_0\, \mathbb{B} \eqend{,}
\label{supertorsiontwobis}
\end{equation}
which implies the undeformed BRST rule for $H$~\eqref{supertorsiontwo}
\begin{equation}
\frac{1}{2} \mathbb{R}_{ab}\, e^a\, e^b = \delta_0 H \eqend{.}
\end{equation}
When we deform $S$, we have at first order in $t$
\begin{equation}
\delta_0 \rightarrow \delta = \delta_0 + t S_1 + \mathcal{O}(t^2)
\end{equation}
with
\begin{equation}
S_1\, \mathbb{B} = - \frac{1}{4} \left( c\, F_- + A_- \, \lambda_- + c\, \lambda_- \right) = - \frac{1}{4} \left( \mathbb{A}_-\, \mathbb{F}_- - A_-\, F_- \right) \eqend{.}
\end{equation}
Hence we obtain
\begin{equation}
\delta\, \mathbb{B} + \mathbb{T} + \frac{t}{4} \mathbb{A}_-\, \mathbb{F}_- = H + \frac{t}{4} A_-\, F_- + \mathcal{O}(t^2)
\label{deltaBdeformed}
\end{equation}
and
\begin{equation}
\frac{1}{2} \mathbb{R}_{ab}\, e^a\, e^b + \frac{t}{4} \mathbb{F}^2_- = \delta \bigl( H + \frac{t}{4} A_-\, F_- \bigr) + \mathcal{O}(t^2) \eqend{.}
\label{supertorsiondeformed}
\end{equation}
Eq.~\eqref{deltaBdeformed} states that, after we have deformed $\delta$, the supertorsion $\mathbb{T}$ ceases to be cohomologous to a 3-form of ghost number 0. However the ``improved'' torsion, which is obtained by adding the super-Chern--Simons invariant $\mathbb{A}\, \mathbb{F}$ to the torsion, is still BRST-equivalent to a 3-form of ghost number 0. Consequently, there exists a super-invariant --- a linear combination of the curvature and YM super-invariants --- which is the $\delta$ variation of a 3-form of ghost number 0, as shown in Eq.~\eqref{supertorsiondeformed}.

\section{Conclusions}
\label{sec:conclusions}

We adopted the BRST framework to discuss supersymmetry anomalies of 4-dimensional $\mathcal{N} = 1$ SQFT's whose $\UR$ symmetry is anomalous. We considered therefore the coupling of SQFT's to $\mathcal{N} = 1$ new minimal supergravity. We pointed out that a supersymmetry BRST cocycle associated to the $\UR$ field exists for this supergravity. This cocycle has a form identical to the chiral supersymmetry anomaly associated to flavor symmetries (different from the R symmetries), once we replace the elementary gaugino with a certain composite of the gravitino which we explicitely computed.

The supersymmetric $\UR$ BRST cocycle is non-trivial in the space of supergravity fields (and ghosts). However, it becomes BRST-exact in the functional space which includes antifields. Equivalently, this cocycle vanishes ``on-shell''. It is therefore removable. However, to remove it ---  precisely because it is not BRST-trivial in the smaller space of fields ---  one needs to deform the supergravity BRST operator $S_0$. This deformation is triggered, at first order in the anomaly coefficient, by a local operator $S_1$ of ghost number 1 which we computed in full detail and which --- to our knowledge --- was not known earlier. We also gave a cohomological, hence intrinsic, characterization of the deformation $S_1$: $S_1$ is the only ghost number 1 local functional derivative that anticommutes with $S_0$, but is not itself an $S_0$-commutator.

For the supersymmetry anomaly to be removable at higher orders in the anomaly coefficient, the deformation triggered by $S_1$ must meet further integrability conditions, which we wrote down. These also can be expressed cohomologically: the cohomology of $S_0$ acting by (anti)-commutators on the space of local functional derivatives must be empty at ghost numbers greater than 1. We did not verify this explicitely, although we expect it to be the case. In any case, since supergravity is not renormalizable, we also expect --- barring some unexpected ``miracle'' --- that deformations of the BRST rules receive non-vanishing contributions at all orders in the anomaly coefficient.

Finally, let us conclude by commenting on the relationship between our results and other recent works on the same topic.

In Ref.~\cite{Papadimitriou:2019yug}, where anomalous supersymmetric Ward identities associated to the $\UR$ anomaly were presented, the question was asked --- in a footnote --- if and how  supersymmetry transformations could consistently be deformed by adding to the transformation of the 2-form field $B_{\mu\nu}$ a term proportional to $c F_{\mu\nu}$. In the same footnote, the author also wonders about the effect of this modification on quantum anomalies. Our Eq.~\eqref{FinalS1} describes precisely such a consistent deformation: as we have shown this removes all the anomalies of the theory, the supersymmetry anomaly together with the bosonic $\UR$ anomaly.

Refs.~\cite{Kuzenko:2019vvi,Bzowski:2020tue} study the issue of the supersymmetry anomaly and of its removal in the language of currents and Wess--Zumino consistency conditions. These works deal both with the supersymmetry anomalies associated to external ``flavor'' symmetries of SQFT's, which we did not discuss here, and with those associated with the R symmetries, the focus of the present paper. The authors of~\cite{Kuzenko:2019vvi,Bzowski:2020tue} analyze the super-algebra of supersymmetry and gauge transformations to derive Wess--Zumino consistency conditions for the various currents. They  emphasize the importance that this superalgebra close off-shell. To this end, they introduce compensator fields belonging to relevant superfields, and for external ``flavor'' symmetries, these compensators belong to vector superfields. On the other hand, the compensators relevant to anomalies related to R symmetries come from superconformal multiplets, as supergravity is obtained in these works by gauge-fixing conformal supergravity. The conclusion of this line of work is that by introducing suitable compensators the anomaly can be ``moved away'' from the supersymmetry sector to the gauge sector.

In the BRST framework, one does not separate ``gauge'' and  ``supersymmetry'' anomalies: they are all captured by a single BRST cocycle, which depends on all ghosts --- both commuting and anticommuting. Of course, different representatives of the same cocycle are possible: ``moving'' the anomaly from one symmetry to another would mean finding a representative of the anomaly cocycle depending on some, but not all, (super-)ghosts, if such a representative exists.

Regarding the anomaly associated to R symmetry --- the focus of the present paper --- it should be kept in mind that a perfectly nilpotent BRST operator acting on the fields of new minimal $\mathcal{N} = 1$ supergravity does exist. Thus, analyzing this theory requires neither superfields nor compensators. We have shown that in this case one can, by deforming the BRST rules, completely remove the anomaly, not just ``move'' it from one symmetry to another. Moreover, this can be done with the fields of the supergravity multiplet alone, without introducing any compensator fields.

It is also worth pointing out that our results hold \emph{at all orders} in the number of gravitinos. In other words, from our deformed BRST rules one can derive one-loop Ward identities involving any number of supercurrents. This is unlike the results of~\cite{Papadimitriou:2019yug,Kuzenko:2019vvi,Bzowski:2020tue}, which seem to be restricted to a fixed number of gravitinos or supercurrents, at least with the explicitly given expressions. Of course, at higher loop orders one needs to determine higher-order deformations of the BRST operator as we have explained.

In this paper, we have not specifically discussed supersymmetry anomalies associated to external ``flavor'' symmetries. We leave a BRST analysis of the issue raised by~\cite{Papadimitriou:2017kzw, Papadimitriou:2019yug, Katsianis:2019hhg, Katsianis:2020hzd, An:2019zok} in this context to the future. Nonetheless, let us briefly comment on this case as well. A BRST formulation of super-Yang--Mills theory coupled to new minimal supergravity does exist, and this formulation includes only the fields of the super-Yang--Mills multiplet in the Wess--Zumino gauge. The equivariant BRST algebra closes without the need for any compensators or antifields, see for example~\cite{Brandt:1996au}.\footnote{To explain this fact in our language, we note that the equivariant $S$ squares to gauge transformations: this is the basic reason why there is no need for compensating gauge transformations in this framework.} The crucial step to obtain this closure is to include a supersymmetry transformation in the BRST transformation rules for the gauge ghost, see our Eq.~\eqref{scYM}. In this formulation, the chiral superanomaly cocycle relative to the external super-Yang--Mills multiplet is not removable even if one includes antifields. The results of~\cite{Kuzenko:2019vvi,Bzowski:2020tue} suggest that, by adding compensating multiplets, it might be possible to construct a representative for this cocycle that does not involve the supersymmetry ghost. It is not clear to us if this remains true in the more economical equivariant BRST formulation that does not include compensators.

\acknowledgments

The work of C. I. and N. R. was supported in part by the Italian Istituto Nazionale di Fisica Nucleare and by Research Funds, F.R.A. 2019, of the Department of Physics of the University of Genoa.

This work has been funded by the Deutsche Forschungsgemeinschaft (DFG, German Research Foundation) --- project nos. 415803368 and 406116891 within the Research Training Group RTG 2522/1.

\appendix

\section{Relation with the BV formalism}
\label{sec: relation with BV}

One could rephrase the discussion of Section~\ref{sec:evanescent anomalies} in the BV language. Our starting equations~\eqref{s1anomaly}
\begin{equations}
S_0 \, \Gamma_\text{eff}[\phi] + t S_1\, \Gamma_0[\phi] &= 0 \eqend{,} \\
S_0 \, \Gamma_0[\phi] &= 0
\end{equations}
can be written as
\begin{equations}
S_0\, \Gamma^\text{BV}_0[\phi, \phi^*] &= 0 \eqend{,} \\
S_0\, \Gamma^\text{BV}_1[\phi, \phi^*] &= 0
\end{equations}
for the antifield-dependent BV actions
\begin{equations}
\Gamma^\text{BV}_0[\phi, \phi^*] &= \Gamma_0[\phi] + \sum_i \phi_i^*\, S_0\, \phi^i \eqend{,} \\
\Gamma^\text{BV}_1[\phi, \phi^*] &= \Gamma_1[\phi] + \sum_i \phi_i^*\, S_1\, \phi^i \eqend{.}
\end{equations}
Therefore, $\Gamma^\text{BV}_1[\phi, \phi^*]$ is a ghost number 0 element of the cohomology of $S_0$ acting on the ``big'' space of both fields and antifields. It is well known~\cite{Barnich:1993vg} that \emph{local} elements of this cohomology as associated to \emph{deformations} of the original classical action $\Gamma_0$. For new minimal $\mathcal{N} = 1$ supergravity, it has been shown~\cite{Brandt:1996au} that the deformations that change the symmetry transformations always involve other external supermultiplets in addition to the supergravity one. However, the BV cohomology problem which is relevant for the removal of anomalies, and which is equivalent to the problem that we solved in the BRST framework, is different from the one considered in~\cite{Brandt:1996au}: in our context $\Gamma_1[\phi]$ is a \emph{non-local} functional of the fields. Hence the BV local cohomology of ghost number 0 computed in~\cite{Brandt:1996au} is not related to deformations of the BRST operator $S_t$ which remove the ``evanescent'' anomaly, and which we computed in this paper.

\section{Details on the computation with FieldsX}
\label{sec:fieldsx}

We include with the paper a Mathematica notebook containing the computations that we did using \textsc{FieldsX}. The notebook is heavily commented and should be mostly self-explanatory: in section 1 of the notebook, the \textsc{FieldsX} package and other \textsc{xAct} packages are loaded. In section 2, we define the manifold, the fields of $\mathcal{N} = 1$ new minimal supergravity, and the Lorentz-covariant derivatives $D$~\eqref{deflorentzd} and $D^\pm$ that are needed, as well as some helper functions to convert between different covariant derivatives and combinations of fields~\eqref{summaryN1newmincombinations}. In section 3, we give the equivariant BRST transformations~\eqref{summaryN1newmin}, the Bianchi identities for the Riemann tensor of the spin connection $\omega_\mu{}^{ab}$ with the torsion expressed using the gravitino (the ghost number 0 component of Eq.~\eqref{generalizedtorsion}), and the vector constraint obtained by acting with the exterior differential on Eq.~\eqref{Hdef}. In section 4, we verify the equivariant BRST algebra~\eqref{BRSTequivariantSone} on the basic fields $\zeta$, $e_\mu{}^a$, $\psi_\mu$, $B_{\mu\nu}$ and $A^-_\nu$, and in section 5 we verify the BRST transformations of the combinations $\omega^-_\mu{}^{ab}$ and $H_\mu$. Section 6 verfies the BRST invariance of the action~\eqref{sugralagrangian} and that the spin connection satisfies its equation of motion~\eqref{spinconnectioneom}, and in section 7 we verify the invariance of the transformation parameters $\gamma^\mu$~\eqref{gammaconsistency}, the Lorentz and $\mathrm{U}(1)$ components of $\phi$ and the $B$ field gauge parameter~\eqref{Bconsistency}.

The anomaly density~\eqref{A4anomalyN1} is shown to fulfill the relation~\eqref{Sanomalyinvariance} in section 8, where we also define the spinorial and topological gaugino and their BRST transformations, as well as some helper functions to ensure that $\d F^- = \d^2 A^- = 0$. Note that compared to the normalization that we use in the paper, the anomaly density in the Mathematica notebook has an extra factor of $(-4)$, which also rescales the first-order deformation $S_1$~\eqref{FinalS1} by the same factor, to ensure that the anomaly is cancelled~\eqref{anomalyfromdeformation}. In the last section 9, we then define the deformation $S_1$ as well as the $\mathrm{U}(1)$ gauge transformation part of $\delta^\text{\tiny YM}_c$, and verify the deformed BRST algebra~\eqref{s0s1consistencytris} on all the basic fields $\zeta$, $e_\mu{}^a$, $\psi_\mu$, $B_{\mu\nu}$ and $A^-_\nu$. Note that also the normalization of the vector $\Theta_\mu$~\eqref{Thetavectordef} differs by a factor $(-4)$ between the paper and the Mathematica notebook.

\bibliographystyle{JHEP}
\bibliography{ir}

\end{document}